\documentclass[aps,pra,twocolumn,showpacs,groupedaddress]{revtex4-1}

\usepackage{amsthm}
\usepackage[english]{babel}
\usepackage[T1]{fontenc}
\usepackage[latin1]{inputenc}
\usepackage{graphicx}
\usepackage{amssymb}
\usepackage{color}
\usepackage{hyperref}
\usepackage{amsmath}

\def\brho {\mbox{\boldmath $\rho$}}

\begin{document}

\title{Universality of Fragmentation in the Schr\"odinger Dynamics 
of Bosonic Josephson Junctions}

\author{Kaspar Sakmann$^{1}$\footnote{E-mail: sakmann@stanford.edu},
Alexej I. Streltsov$^{2}$,
Ofir. E. Alon$^{3}$,
and Lorenz S. Cederbaum$^{2}$}

\affiliation{$^1$ 
Department of Physics, Stanford University, Stanford, California 94305, USA}

\affiliation{$^2$ Theoretische Chemie, Physikalisch-Chemisches Institut, 
Universit\"at Heidelberg,\\
Im Neuenheimer Feld 229, D-69120 Heidelberg, Germany}

\affiliation{$^3$ Department of Physics, University of Haifa at Oranim, 
Tivon 36006, Israel}

\date{\today}

\begin{abstract}
The many-body Schr\"odinger dynamics of a one-dimensional bosonic Josephson junction is investigated
for up to ten thousand bosons and long times.
The initial states are fully condensed and the interaction strength is weak.
We report on a universal fragmentation dynamics on the many-body level:  
systems consisting of different numbers of particles fragment to the same value 
at constant mean-field interaction strength.
The phenomenon manifests itself in observables such as the correlation functions of the system. 
We explain this universal fragmentation 
dynamics analytically based on the Bose-Hubbard model. 
We thereby show that the extent to which many-body effects become important at  
later times depends crucially on the initial state. 
Even for arbitrarily large particle numbers and arbitrarily weak interaction strength
the dynamics is many-body in nature and the fragmentation universal. 
There is no weakly interacting limit where the Gross-Pitaevskii mean-field is valid for long times.
\end{abstract}

\maketitle

\section{Introduction}
Over the last few years ultracold bosons in double-well potentials have
allowed the exploration of quantum many-body dynamics in a 
highly controllable manner. Among the phenomena predicted or observed 
are tunneling and self-trapping \cite{Milburn97,Smerzi97, e1,e2,Zibold10}, 
Josephson oscillations \cite{Smerzi97,e1,e2,Zibold10}, collapse and revival sequences \cite{Milburn97}, 
squeezing \cite{Kasevich0}, and matter wave interferometry \cite{Schumm05},
to name a few.
While tunneling, self-trapping and Josephson oscillations have explanations 
on the mean-field level, other phenomena, e.g., collapse and revival sequences 
and fragmentation dynamics require many-body treatments such as the Bose-Hubbard model \cite{Milburn97}
or even the full many-body Schr\"odinger equation \cite{BJJexact,BJJexact2}.  
Of particular relevance to this work is condensate fragmentation \cite{NoS82,LeggettBook}, 
i.e. the occurrence of several macroscopically occupied quantum states as opposed 
to conventional Bose-Einstein condensation with just one macroscopically occupied state
\cite{PeO56}.
Condensate fragmentation has been predicted to occur \cite{Spekkens99,MCHB,ueda,Fischer09,BJJexact}
and observed in the ground state \cite{JoergNatPhys06},  see also
\cite{Kasevich0} in this context. It is a many-boson phenomenon due to interparticle interactions.
Generally, stronger interactions lead to more fragmentation, 
but the question to what extent a condensate fragments usually depends intricately 
on the number of bosons and the interaction strength. 
 
In bosonic Josephson junctions self-trapping 
-- the inhibition of tunneling due to interparticle interactions --
occurs from a critical value of the interaction strength onwards within two-mode Gross-Pitaevskii mean-field theory  \cite{Milburn97,Smerzi97}.
This critical value provides a reference point for the importance of interactions in the system. 
Two-mode GP mean-field theory was successfully applied in the description 
of bosonic Josephson junction experiments at short time scales \cite{e1,e2,Zibold10}. However,  
for longer times the dynamics can leave the realm of mean-field theory and 
make a many-body description necessary, 
see, e.g., Refs. \cite{Milburn97,Raghavan99,Vardi01,BJJexact}. 

This paper contains the following main results. Firstly, we investigate the long-time many-body dynamics 
of large Bose-Einstein condensates (BECs) of up to ten thousand bosons.  
To this end we solve the time-dependent many-body Schr\"odinger
equation numerically for initially fully condensed BECs in a one-dimensional bosonic Josephson junction. 
We consider interaction strengths that are so weak that self-trapping cannot occur. 
We report on the existence of a universal many-body
fragmentation dynamics in bosonic Josephson junctions:
systems consisting of different numbers of bosons
all fragment to the same value at fixed mean-field interaction strength.
The occurrence of such a universal many-body dynamics is unexpected and is 
shown here explicitly based on the many-body Schr\"odinger equation for up to 10000 bosons. 
The phenomenon manifests itself in observables, such as the correlation functions.

Secondly, using the Bose-Hubbard (BH) model we explain analytically how 
the universal fragmentation dynamics depends on the initial state and its occurrence
for arbitrarily large numbers of particles and arbitrarily weak mean-field interaction strength. 
We show that for a whole class of condensed initial states the analytical calculations allow for precise 
predictions about the fragmentation of the BEC 
after the collapse of the density oscillations and thus about  the extent 
to which the many-body dynamics remains within the realm of mean-field.
We thereby show that there is no limit for which GP theory remains valid at long times.

\section{Theoretical Framework}
In this section we introduce the theoretical methods and quantities  
used in this work to explore the dynamics 
of a one-dimensional bosonic Josephson junction.
In the following we suppress the time argument whenever possible.

\subsection{Many-body Hamiltonian, wavefunction and parameters}
It is convenient to use dimensionless units defined
by dividing the Hamiltonian by $\frac{\hbar^2}{mL^2}$,
where $m$ is the mass of a boson and $L$ is a length scale.
The full many-body Hamiltonian then reads 
\begin{equation}\label{HAM}
H=\sum_{i=1}^N h(x_i) + \sum_{i<j}W(x_i-x_j),
\end{equation}
where $h(x)=-\frac{1}{2}\frac{\partial^2}{\partial x^2} + V(x)$ 
with a trapping potential $V(x)$ and an 
interparticle interaction potential 
$W(x-x')=\lambda_0\delta(x-x')$.
For the potential $V(x)$ we choose a double-well constructed by connecting 
two harmonic potentials 
\begin{equation}
V_{\pm}(x)=\frac{1}{2}(x\pm 2)^2
\end{equation}
with a natural cubic spline at $|x|=0.5$.  This uniquely defines the external potential $V(x)$. 
The  even-symmetry ground state  $\phi_g$ and odd-symmetry excited state  $\phi_u$ of the one-particle 
double-well Hamiltonian $h(x)$ allow to define several commonly used parameters. 
First two left- and right-localized Wannier functions 
\begin{equation}\label{Wannier}
\phi_{L,R}=\frac{1}{\sqrt{2}}(\phi_g\pm\phi_u)
\end{equation}
are constructed. This defines the hopping parameter 
\begin{equation}
J= -\langle \phi_L \vert h \vert \phi_R\rangle=2.2334\times10^{-2},
\end{equation}
as well as the interaction parameter
\begin{equation}\label{defU}
U=\lambda_0\int{dx |\phi_L(x)|^4}
\end{equation}
and the single particle tunneling time scale  $t_{Rabi}=\pi/J=140.66$.

In second quantization the Hamiltonian (\ref{HAM}) of the many-boson system reads
\begin{equation}\label{MBHam}
 \hat H = \int dx \, \hat\Psi^\dag(x) h(x) \hat\Psi(x) + \frac{\lambda_0}{2} \int dx \, \hat\Psi^\dag(x) \hat\Psi^\dag(x) \hat\Psi(x) \hat\Psi(x).
\end{equation}
The theoretical approaches employed in this work,
namely the Bose-Hubbard model (dimer) and the multiconfigurational time-dependent Hartree for bosons (MCTDHB) method \cite{MCTDHB}
can be derived from different truncations of the field operator
\begin{equation}\label{FieldOP}
\hat\Psi(x) = \sum_j \hat b_j \phi_j(x).
\end{equation}
Here, $\{\phi_j(x)\}$ denotes a complete orthonormal set of 
orbitals, $\hat b_j$ annihilates a boson in $\phi_j(x)$ and 
$[\hat b_i,\hat b_j^\dagger]=\delta_{ij}$. In practice, the infinite set $\{\phi_j(x)\}$ is 
truncated to a finite number $M$.
Associated with a truncation of the field operator to $M$ terms
is an expansion of the $N$-boson wavefunction
\begin{equation}\label{PSI}
\vert\Psi\rangle = \sum_{\vec n} C_{\vec n}|\vec n\rangle,
\end{equation}
with expansion coefficients $C_{\vec n}$ and basis vectors
\begin{equation}
\vert \vec n \rangle = \frac{\hat b_1^{\dagger n_1}\hat b_2^{\dagger n_2} \cdots\hat b_M^{\dagger n_M}}{\sqrt{n_1!n_2!\cdots n_M!}}\vert 0 \rangle.
\end{equation}
Here, $n_1+\dots+n_M=N$ is implied. There are $\binom{N+M-1}{N}$ expansion coefficients $C_{\vec n}$.

\subsection{Bose-Einstein condensation and fragmentation}\label{BECandFrag}
For a given many-boson wavefunction $\vert\Psi\rangle$ all one-body observables can be 
computed from the  first-order reduced density matrix
\begin{align}\label{RDM}
\rho^{(1)}(x\vert x') &= \langle \Psi \vert \hat \Psi^\dagger(x')\hat \Psi(x) \vert \Psi \rangle\nonumber \\
&=\sum_{ij}\rho_{ij}\phi_j(x)\phi_i^\ast(x')\nonumber\\
&=\sum_i n^{(1)}_i
\alpha_i^{(1)}(x)\alpha_i^{(1)\ast}(x'),
\end{align} where $\rho_{ij}=\langle\Psi\vert\hat b_i^\dagger\hat b_j\vert\Psi\rangle$ and the eigenfunctions $\alpha_i^{(1)}(x)$ and eigenvalues $n^{(1)}_1\ge n^{(1)}_2\ge\dots$  
are known as natural orbitals and natural occupations, 
respectively. 
The natural occupations satisfy $\sum_{i} n^{(1)}_i = N$ 
and the single-particle density is given by 
\begin{equation}
\rho(x)=\rho^{(1)}(x\vert x).
\end{equation}
Bose-Einstein condensation is defined as follows: 
if only one nonzero eigenvalue $n^{(1)}_1={\mathcal O}(N)$ exists, the system is condensed \cite{PeO56}. 
If there is more than one such eigenvalue, the BEC
is said to be fragmented \cite{NoS82,Spekkens99,MCHB,ueda,Sak08}.

For our purposes it is convenient to quantify fragmentation using the quantity 
\begin{equation}\label{defFrag}
f=\frac{1}{N}\sum_{i>1} n^{(1)}_i
\end{equation}
which we refer to as the fragmentation for simplicity.
For fully condensed states, i.e. $n_1^{(1)}=N$, one finds  $f=0$, whereas $0\le f<1$ in general. 
Truncating the field operator to $M$ terms leads to an $M$ by $M$
first-order reduced density matrix $\rho_{ij}$, implying $f\le\frac{M-1}{M}$.
Fragmentation manifests itself in 
correlation functions and fringe visibilities \cite{Sak08}. 
Several experiments have measured the effects of fragmentation, 
see for instance \cite{Kasevich0,JoergNatPhys06,Westbrook07}. 

In the strict sense the definition of BEC given above  makes use of the 
thermodynamic limit \cite{PeO56}. It is nevertheless common practice 
to apply it also to finite, trapped systems, such as those realized in experiments \cite{LeggettBook}.

In the thermodynamic limit Bose-Einstein condensation has been interpreted as
the largest eigenvalue of the one-particle reduced 
density matrix, $n^{(1)}_1$ being an extensive rather than an intensive quantity \cite{PeO56}.
The fragmentation $f$ as defined in Eq. (\ref{defFrag}) would then be an intensive quantity.
This interpretation does not hold for trapped, finite systems: 
the ground states of some trapped, interacting BECs of finite size are known to 
be fragmented, i.e. $f>0$, see, e.g., \cite{Spekkens99,MCHB,ueda,Fischer09, JoergNatPhys06, Kasevich0}.
However, it can be rigorously proven that $f=0$  for the ground states of trapped BECs 
at any finite mean-field interaction strength $\lambda_0(N-1)$ in 
the limit $N\rightarrow\infty$ \cite{LiebSeiringer2002}.
Thus, fragmentation is dependent on $N$ and cannot be interpreted as
an intensive quantity in trapped finite systems.
In nonequilibrium dynamics problems, which we discuss in this work, 
the fragmentation generally does not only change with the particle 
number and mean-field interaction strength, but also in time.

\subsection{Probability in the left well and correlation functions}
In the dynamics of Bose-Einstein condensates in double-well potentials
it is instructive to consider the integral over the probability density in the left half of the double-well potential
\begin{equation}\label{pL}
p_L(t)=\frac{1}{N}\int_{-\infty} ^0 \rho(x;t) dx.
\end{equation}
The probability in the right half of space  then follows from normalization $p_R(t)=1-p_L(t)$. 

Correlation functions can be derived from the reduced density matrices of all orders, e.g. 
the single-particle momentum distribution and the single-particle density 
derive from the first-order reduced density matrix, Eq (\ref{RDM}).
It is possible to answer the question whether a condensate is fragmented or not 
from the simultaneous measurement of the single-particle density and momentum distribution. 
 
However, second-order correlation functions are particularly appealing because they 
provide the answer to this question through the measurement of a single quantity.  
Second-order correlation functions of Bose-Einstein condensate 
have been measured in many recent experiments see, e.g.,  Refs.  \cite{Bloch05,Westbrook07,Schmiedmayer12}.

The momentum correlation function is defined as
\begin{equation}
g^{(2)}(k,k')=\frac{\rho(k,k')}{\rho(k)\rho(k')},
\end{equation} 
where 
\begin{equation}
\rho(k)=\langle\Psi\vert \hat{\Psi}^\dag(k) \hat{\Psi}(k)\vert \Psi\rangle
\end{equation}
is the single-particle 
and \begin{equation}
\rho(k,k')=\langle\Psi\vert \hat{\Psi}^\dag(k)\hat{\Psi}^\dag(k') 
\hat{\Psi}(k') \hat{\Psi}(k)\vert \Psi\rangle
\end{equation}  the two-particle momentum distribution.
For fully condensed states the momentum correlation function is constant, $g^{(2)}(k,k')=1-1/N$,
indicating that there are no correlations between momenta, 
see, e.g., \cite{Gla99,Sak08}.  
The presence of momentum correlations in BECs 
therefore indicates unambiguously the presence of fragmentation.

\subsection{Bose-Hubbard model}
A popular many-body model for the description of bosons in a symmetric double-well is the BH model for two sites. 
The BH Hamiltonian for a double-well potential can be obtained explicitly from Eq. (\ref{MBHam}) by
restricting the field operator to a sum of only two terms
\begin{equation}\label{BHfieldop}
\hat\Psi(x) = \hat b_L \phi_L(x) + \hat b_R \phi_R(x).
\end{equation} 
using the left- and right-localized Wannier functions, Eq. (\ref{Wannier}).

By substituting Eq. (\ref{BHfieldop}) into
the many-body Hamiltonian (\ref{MBHam}), 
neglecting the off-diagonal interaction terms (e.g., ~$\hat b^\dag_L \hat b^\dag_L \hat b_L \hat b_R$), 
and by eliminating the diagonal one-body terms (e.g., ~$\hat b^\dag_R \hat b_R$), 
one readily arrives at the Bose-Hubbard Hamiltonian
\begin{equation}\label{BHHam}
\hat H_{BH} = -J \left(\hat b_L^\dag \hat b_R + \hat b_R^\dag \hat b_L\right) + \frac{U}{2}
\left(\hat b_L^\dag \hat b_L^\dag \hat b_L \hat b_L + 
\hat b_R^\dag \hat b_R^\dag \hat b_R \hat b_R \right).
\end{equation}
The Bose-Hubbard Hamiltonian describes the evolution of the many-particle 
system in the space restricted to two localized
time-independent modes.
The Bose-Hubbard time-dependent
$N$-boson wavefunction takes on the form
\begin{equation}
|\Psi_{BH}(t)\rangle = \sum_{n_L=0}^N C_{n_L}(t) |n_L,N-n_L\rangle,
\end{equation}
where the time-dependent coefficients $\{C_{n_L}(t)\}$ governing 
the evolution of the junction are simply obtained from the first-order equation
\begin{equation}
\hat H_{BH} |\Psi_{BH}(t)\rangle = i |\dot\Psi_{BH}(t)\rangle.
\end{equation}
$\{|n_L,N-n_L\rangle\}$ are the Fock states assembled
from the two `modes' $\hat b^\dag_L$ and $\hat b^\dag_R$.
Within the BH model the probability in the left well, Eq. (\ref{pL}), is conventionally identified with the occupation of the orbital $\phi_L$
\begin{equation}
p_L(t)=\frac{1}{N}\langle \Psi_{BH}(t) \vert \hat b_L^\dag \hat b_L\vert \Psi_{BH}(t)\rangle.
\end{equation}
The field operator contains only $M=2$ terms in the BH model and thus 
the fragmentation of the BEC reduces here to the
occupation of the second natural orbital, $f=n_2^{(1)}/N\le50\%$.
Note that in the full many-body Schr\"odinger dynamics of bosonic 
Josephson junctions more than two natural orbitals can become 
occupied \cite{BJJexact}.

The dynamics governed by the Bose-Hubbard Hamiltonian 
occurs solely on the lowest band of the one-particle double-well problem.
This emanates from Eq. (\ref{BHfieldop}),
namely from retaining the first two terms only in
the expansion of the field operator.
It is in principle possible to retain more terms and include also higher 
band Wannier functions.
This results in multi-band Bose-Hubbard models
which approach the full many-particle Hamiltonian (\ref{MBHam}) with increasing number of included bands.
However, in general many Wannier functions (Bloch bands)
are needed to faithfully represent the full many-particle Hamiltonian (\ref{MBHam})
particularly when time-dependent phenomena are investigated.

\subsection{Multiconfigurational time-dependent Hartree for bosons (MCTDHB)}
In order to compute the time-evolution of the many-body Schr\"odinger equation 
we use the  multiconfigurational time-dependent Hartree for bosons (MCTDHB) method, 
see Ref. \cite{MCTDHB} for theory and implementation  
and Refs.  \cite{BJJexact,BJJexact2,optimal,triplewell,HIM} for applications.
In short, the MCTDHB method employs variational time-adaptive orbitals in the expansion 
of the many-boson wavefunction. With increasing number of orbitals the method converges to the exact many-body 
Schr\"odinger result \cite{BJJexact,BJJexact2,triplewell}, as was also 
shown by comparison to an exactly-solvable many-body model \cite{HIM}. 

A faithful representation of the full many-body Hamiltonian (\ref{MBHam}) using time-independent orbitals 
often requires truncating the field operator (\ref{FieldOP}) after a large number of terms. 
Even for few bosons this leads to prohibitively large Fock spaces.
In the MCTDHB method time-dependent orbitals are used as opposed to the time-independent ones 
in the expansion of the field operator
\begin{equation}\label{MCTDHBFieldOP}
\hat\Psi(x) = \sum_{j=1}^M \hat b_j(t) \phi_j(x,t).
\end{equation}
The orbitals are determined from the time-dependent variational principle. 
We briefly review the key features of the MCTDHB method.

Consider $M$ time-adaptive orbitals. When distributing $N$ bosons
over the $M$ orbitals the resulting time-dependent many-boson wavefunction reads:
\begin{equation}\label{PSIMCTDHB}
\vert\Psi(t)\rangle = \sum_{\vec n} C_{\vec n}(t)  |\vec n;t\rangle.
\end{equation}
In Eq. (\ref{PSIMCTDHB}) the $\{C_{\vec n}(t)\}$ are the time-dependent expansion coefficients
and $\{|\vec n;t\rangle\}$ the time-dependent permanents (symmetrized Hartree products)
built from the $M$ time-dependent orbitals $\{\phi_j(x,t)\}$. 

In the MCTDHB method the coefficients $\{C_{\vec n}(t)\}$ and the orbitals $\{\phi_j(x,t)\}$ 
of the many-boson wavefunction (\ref{PSIMCTDHB}) are determined by the time-dependent 
variational principle.
Explicitly, the MCTDHB equations of motion 
are derived by requiring stationarity of the many-body Schr\"odinger action functional 
\begin{align}\label{actionfunc}
 S[\{C_{\vec n}(t)\},\{\phi_j(x,t)\}] = \int dt \biggl\{\langle\Psi(t)|\hat H-i\frac{\partial}{\partial t}|\Psi(t)\rangle \nonumber \\
- \sum_{k,j=1}^{M} \mu_{kj}(t)[\langle\phi_k|\phi_j\rangle - \delta_{kj}]\biggr\}
\end{align}
with respect to variations of the time-dependent expansion coefficients
$\{C_{\vec n}(t)\}$ and orbitals $\{\phi_j(x,t)\}$.
Here, $\{\mu_{kj}(t)\}$ are time-dependent Lagrange multipliers 
which ensure the
orthonormality of the orbitals throughout their propagation in time.
The derivation is quite lengthy but otherwise straightforward,
see \cite{MCTDHB} for details.
Here we quote the final result for the MCTDHB equations of motion for the interaction potential $W(x-x')=\lambda_0\delta(x-x')$ which read:
\begin{equation}\label{phidot}
 i|\dot\phi_j\rangle = \hat{\mathbf P} \left[h |\phi_j\rangle + \lambda_0 \sum_{k,s,l,q=1}^M \{\brho(t)\}^{-1}_{jk}  
 \rho_{kslq} \phi_s^\ast \phi_l |\phi_q\rangle \right], 
\end{equation}
with $j=1,\ldots,M$ for the time-dependent orbitals 
and
\begin{equation}\label{S10}
 \mathbf{H}(t) \mathbf{C}(t) = i \dot{\mathbf{C}}(t), \qquad H_{\vec n\vec n'}(t) = \langle\vec n;t|\hat H|\vec n';t\rangle
\end{equation}
for the expansion coefficients.
Here, $\hat{\mathbf P} = 1 - \sum_{u=1}^M |\phi_u\rangle\langle\phi_u|$ is a projector operator
on the complementary space of the time-adaptive orbitals, 
$\brho(t)$ the reduced one-particle density matrix,
$\{\rho_{kslq}\}$ the elements of the two-particle reduced density matrix, 
and $\mathbf{C}(t) = \{C_{\vec n}(t)\}$ the vector of expansion coefficients.

The use of $M$ optimized time-dependent orbitals leads to much faster numerical convergence 
to the full many-body Schr\"odinger results than an expansion in $M$ time-independent orbitals.
Thereby, problems involving large numbers of bosons 
can be solved on the full-many-body Schr\"odinger level.
Indeed, numerical convergence
to the exact time-dependent solution
of the many-boson Schr\"odinger equation 
in closed \cite{BJJexact} and open \cite{PNAS}
trap potentials has been reported, 
as well as benchmarking against an 
exactly-solvable
many-boson model \cite{HIM}.
For more details see \cite{MCTDHB}.
The MCTDHB method has been cast as an efficient algorithm into a software package \cite{Package}.
In practice, Eq. (\ref{S10}) is simplified by a mapping of the configuration space \cite{mapping}
which enables the efficient handling of millions of time-adaptive permanents \cite{Package}.

\subsection{Gross-Pitaevskii mean-field}\label{GPmeanfield}
One particular limiting case of the MCTDHB method deserves special attention:
namely  the case where only a single time-dependent orbital is used in the expansion of the field operator Eq. (\ref{MCTDHBFieldOP}) and the many-body wavefunction Eq. (\ref{PSIMCTDHB}), 
i.e. when $M=1$. In this case all particles occupy the same time-dependent orbital $\phi(x,t)$ at all times, 
i.e. there is only ever a single coefficient $C_{\vec n}$ and the solution of Eq. (\ref{S10}) becomes trivial. 
Thus, only Eq. (\ref{phidot})  needs to be solved which differs from the well-known Gross-Pitaevskii mean-field equation 
\begin{equation}\label{GPequation}
i\dot\phi(x,t) = [h(x)+\lambda_0(N-1) \vert \phi(x,t)\vert^2] \phi(x,t).
\end{equation}
only by an irrelevant overall phase. In other words, for $M=1$ the MCTDHB method reduces to GP mean-field theory.
In GP theory the only free parameter is 
\begin{equation}\label{lambda}
\lambda=\lambda_0(N-1)
\end{equation} 
which is known as the mean-field interaction strength. 
In GP mean-field theory systems consisting of different numbers of particles, but with the same value of $\lambda$ lead to identical dynamics.
Note that increasing  $N$ at constant $\lambda$ implies  a decreasing interaction strength $\lambda_0$.

By approximating the time-dependent Gross-Pitaevskii orbital  as
\begin{equation}
\phi(x,t)=a_L(t)\phi_L(x)+a_R(t)\phi_R(x)
\end{equation}
with $\vert a_L\vert^2+\vert a_R\vert^2=1$ a two-mode GP mean-field model 
can be derived  \cite{Milburn97,Smerzi97}. The model maps the dynamics of the BEC
onto that of a classical nonrigid pendulum.
Within the two-mode GP mean-field model the parameter
\begin{equation}\label{defLambda}
\Lambda=\frac{U(N-1)}{2J}
\end{equation}
plays an important role. Self-trapping, i.e. the inhibition of tunneling becomes possible 
if $\Lambda\ge\Lambda_c=2$ \cite{Milburn97,Smerzi97}. 
The parameter $\Lambda$ is proportional to the mean-field interaction strength 
$\lambda$, see Eq. (\ref{defU}). 

\section{Results}
In this section we show that there is an unexpected universal dynamics on the full many-body Schr\"odinger 
level. We explain this universal many-body dynamics analytically and numerically using the BH model.

\subsection{Many-body Schr\"odinger dynamics}
In the following we study the many-body Schr\"odinger dynamics in the bosonic Josephson junction  
for different particle numbers.
We choose fully condensed initial states which are the
GP ground states of the potential $V_+(x)$. 
Thus, initially the BECs are located in the left well.
For each particle number $N$  the interaction strength $\lambda_0$ is chosen such that
the mean-field interaction strength $\lambda$, defined in Eq. (\ref{lambda}), is constant. 
We begin with $\lambda=0.152$ which  corresponds 
to $\Lambda=1.33$, well below the critical value $\Lambda_c=2$ 
for self-trapping \cite{Milburn97,Smerzi97}.

 \begin{figure}
 \includegraphics[scale=0.4,angle=0]{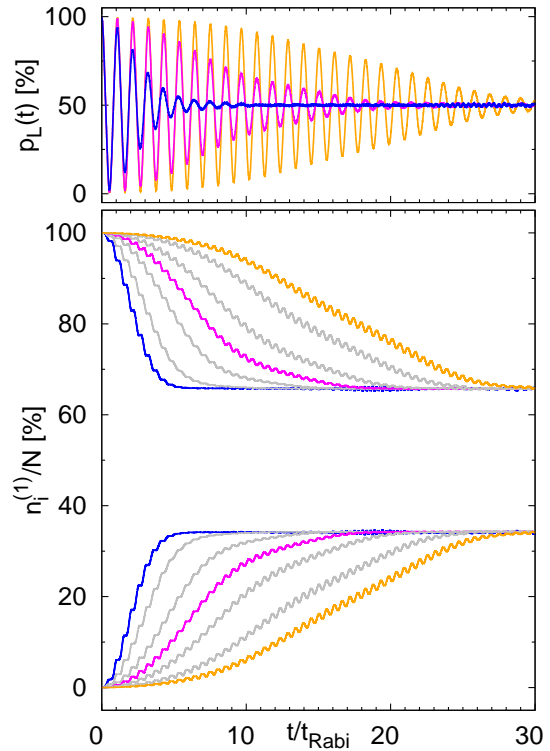}
 \caption{(color online): Universality of the many-body Schr\"odinger fragmentation dynamics. 
 Top:  shown is the probability in the left potential well $p_L(t)$ as a function of time for different numbers of particles.   
 For all particle numbers the initial state is fully condensed, all bosons are in the left well $p_L(0)=1.0$, 
 and the mean-field interaction strength is $\lambda=0.152$ ($\Lambda=1.33$). 
 As the BECs tunnel back and forth through the barrier, the density oscillations collapse.  
 Particle numbers: N=100 (blue), 1000 (magenta), 10000 (orange).
 Bottom: shown are the corresponding natural occupations $n_i^{(1)}/N$ as a 
 function of time.   The grey lines represent
 particle numbers in between: N=200, 500, 2000 and 5000.
 Only two natural occupations $n_1^{(1)}$ and $n_2^{(1)}$ become significantly 
 occupied in the dynamics. Thus, the fragmentation $f$ is given here by $f\approx n_2^{(1)}/N$, see Eq. (\ref{defFrag}).
 The fragmentation reaches a plateau at $f=f_{col}=34\%$ after 
 the collapse of the density oscillations for all particle numbers.
 The nature of the BEC changes from condensed to fragmented.
 All quantities shown are dimensionless.}
 \label{F1}
 \end{figure}

Fig. \ref{F1} (top) shows the probability in the left well
$p_L(t)$ as a function of time for $N=100-10000$ bosons. 
The density tunnels back and forth through the potential 
barrier. However, the amplitude of the oscillations decreases and 
eventually the oscillations collapse.  
With increasing particle number the collapse occurs at later times.  
It is well known that the collapse is present in the many-body dynamics, 
but not within GP mean-field theory \cite{Milburn97,BJJexact}.
We will therefore investigate the many-body nature of the BEC
during the collapse in more detail.

In Fig. \ref{F1} (bottom)  the corresponding natural 
occupations are shown.
Since the many-body wavefunction is initially condensed,
there is only one natural occupation $n^{(1)}_1=N$ at $t=0$
and the fragmentation $f$ is zero, see Eq. (\ref{defFrag}).  
However, as the condensate tunnels back and forth through the barrier a second natural orbital becomes
occupied. 
The nature of the BEC changes from condensed to fragmented
as the density oscillations collapse. 
We stress that these results represent the 
many-body Schr\"odinger dynamics of the Hamiltonian (\ref{HAM}).

For the smallest particle number,  $N=100$ bosons,  $M=4$ orbitals were used in the MCTDHB computation 
and the result is numerically exact. However, only $M=2$ time-dependent variationally optimized orbitals 
are needed here: there is only a small quantitative difference between computations with $M=4$ and $M=2$ orbitals.
Until time $t=30t_{Rabi}$ the occupations of the third and fourth natural orbitals together stay 
below $0.1\%$ and are not shown in Fig. \ref{F1}. 
Therefore, $M=2$ orbitals were used in all computations for larger particle numbers.

From Fig. \ref{F1} (bottom) we see that the fragmentation dynamics is universal in the following sense:
at constant mean-field interaction strength $\lambda$ systems consisting 
of {\em different} numbers of bosons fragment to the {\em same} value 
during the collapse of the density oscillations.  
The two largest natural occupations reach plateaus at 
about $n^{(1)}_1/N=66 \%$ and $n^{(1)}_2/N=34\%$. 
Using Eq. (\ref{defFrag}) we find for the fragmentation after 
the collapse of the density oscillations 
$f=34\%$ regardless of the number of particles involved. 
From now on we refer to the fragmentation after  
the collapse of the density oscillations as $f_{col}$. 
 
This universal behavior is fundamentally different from that within GP theory, where
the dynamics is strictly identical for all systems with the same value of $\lambda$, 
regardless of the number of particles, see Sec. \ref{GPmeanfield}. 
The GP mean-field approximation excludes any possibility for fragmentation.
Here, we report on a universal fragmentation dynamics 
which occurs naturally on the many-body level, i.e., 
without being enforced by an approximation. 
Fragmentation of condensates is usually strongly dependent on the number of 
particles in the system \cite{Spekkens99,MCHB,ueda,Fischer09,BJJexact}, see also the discussion of Ref. \cite{LiebSeiringer2002} 
in Sec. \ref{BECandFrag} in this context.
Hence, the appearance of a universal fragmentation dynamics is highly unexpected. 

We note that the times until a given fragmentation is reached increases
only logarithmically with the number of particles, implying that the phenomenon occurs also for large 
particle numbers. Details on this logarithmic scaling are given in appendix \ref{fragtime}. 

\begin{figure}
 \includegraphics[scale=0.35,angle=-90]{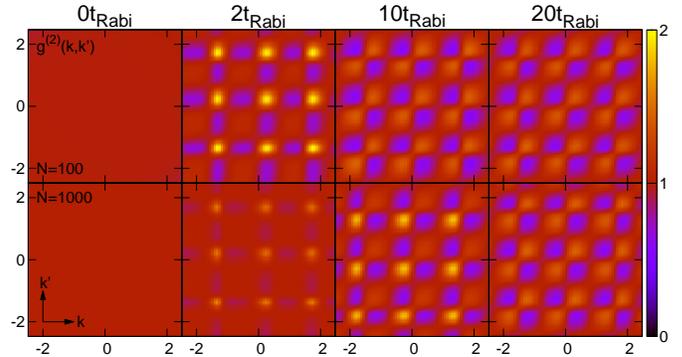}
 \caption{(color online): Universality of the many-body Schr\"odinger two-particle momentum correlations. 
 Shown is the momentum correlation function $g^{(2)}(k,k')$ at different times for two of the cases 
 shown in Fig. \ref{F1}. 
 Top row: $N=100$ bosons. Initially the system is condensed and there are no correlations (leftmost panel).
 During the collapse of the density oscillations correlations build up and the momentum correlation function varies 
 rapidly in time (second panel).
 For $t\ge7t_{Rabi}$ the density oscillations have collapsed, see Fig. \ref{F1} (top). 
 The structure of $g^{(2)}(k,k')$ remains practically constant in time then (third and fourth panel).
 Bottom Row: Same as top row, but for $N=1000$ bosons. The collapse takes until about $t=20t_{Rabi}$ 
 for this larger particle number and the structure of $g^{(2)}(k,k')$ becomes 
 constant from then on (fourth panel). The values of $g^{(2)}(k,k')$ after the 
 collapse are the same for both particle numbers and show how 
 the universality of fragmentation leads to the universality of directly observable quantities.
 All quantities shown are dimensionless.}
 \label{F2} 
 \end{figure}

In order to establish how the universality of fragmentation can be observed, 
we focus on the momentum correlation function $g^{(2)}(k,k')$.
Fig. \ref{F2} shows $g^{(2)}(k,k')$  at different times for $N=100$ and $N=1000$ bosons.
The initial state is condensed $n_1^{(1)}=N$ and therefore $g^{(2)}(k,k')=1-1/N$, i.e. $g^{(2)}(k,k')$ is completely flat. 
As the system starts to fragment, correlations between particles build up and $g^{(2)}(k,k')$ exhibits a time-varying structure. 
However, once the density oscillations have collapsed  $g^{(2)}(k,k')$ is nearly constant in time.  
It is clearly seen from the rightmost panels that the momentum correlation functions for $N=100$ and $N=1000$ bosons 
are the same then. This is the case for all particle numbers considered in this work. 
Thus, the universality of fragmentation manifests itself unambiguously 
as a measurable universality of momentum correlations. 

The periodic spacing of $g^{(2)}(k,k')$ in the rightmost panels is determined by the 
distance $d=4$ between the two wells of the double well potential.
This becomes clear by considering an analytical model for a fragmented condensate consisting of two 
fragments of $N/2$ bosons located in the left and right wells with equal 
occupations $n^{(1)}_1/N= n^{(1)}_2/N =50\%$. 
For such a condensate one finds $g^{(2)}(k,k')\propto 1+N/(N-1)\cos[(k-k')d]/2$, see Ref. \cite{Sak08}, 
i.e. $g^{(2)}(k,k')$ becomes periodic in $k$ and $k'$ with period  $2\pi/d$.
The case shown in Fig. \ref{F2} is more complicated and lies in between the extremes of a fully condensed 
state with a flat  $g^{(2)}(k,k')$ and the fragmented model state described above.
However, the periodicity of $g^{(2)}(k,k')$ in $k$ and $k'$ with period $\pi/2$ is already clearly visible.

\subsection{Bose-Hubbard dynamics}
So far we have established the universality of fragmentation of the many-body Schr\"odinger dynamics.
We will now use a simple model to  show that the universality of fragmentation
is a general many-body phenomenon that exists for a wide range of initial conditions and explain its origin.
As a model we use the two-mode BH Hamiltonian
Eq. (\ref{BHHam}). 

\begin{figure}
 \includegraphics[scale=0.35,angle=-90]{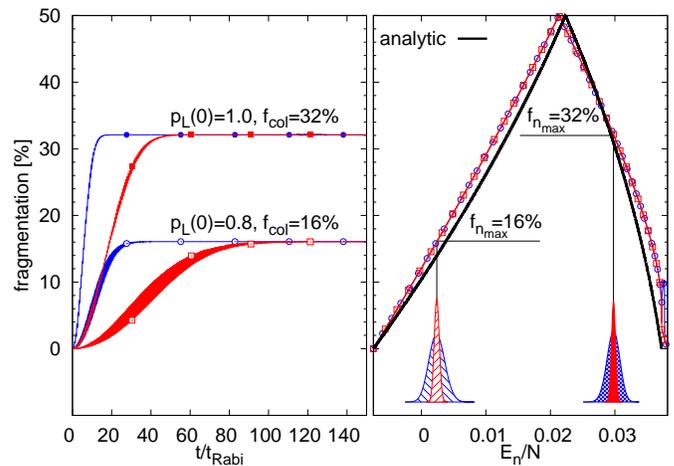}
 \caption{(color online): Universality of the Bose-Hubbard fragmentation dynamics. 
 Left panel: Shown is the fragmentation $f$ as a function of time for different condensed initial 
 states and for different numbers of particles. The initial states $\vert \Psi_0\rangle$ are defined in Eq. (\ref{psi0}).
 The mean-field interaction strength is the same as in Fig. \ref{F1} for all particle numbers. 
 For initial states with $p_L(0)=1.0$ the BH fragmentation reaches a plateau 
 after the collapse of the density oscillations at $f=f_{col}=32\%$, 
 not far from the many-body Schr\"odinger result, see Fig. \ref{F1}.
 The approximative formula for $\Lambda\ll1$, Eq. (\ref{weakfcol}), predicts $f_{col}=1/2-\Lambda/8=33.3\%$, which is also very close. 
 Similarly, for condensed initial states with $p_L(0)=0.8$ a plateau is reached at $f=f_{col}=16\%$.
 Parameter values: $N=1000, p_L(0)=1.0: {\color{blue}\bullet}$;  
 $N=1000, p_L(0)=0.8: {\color{blue}\odot}$; $N=10000, p_L(0)=1.0: {\color{red}\blacksquare}$;
 $N=10000, p_L(0)=0.8: {\color{red}\boxdot}$.
 Right panel: for each eigenstate $\vert E_n\rangle$ of the BH model 
 its fragmentation $f_n$ is shown as a function of the eigenstate energy per particle $E_n/N$. 
 Curves for different particle numbers lie on top of each other 
 (N=1000: ${\color{blue} \odot}$; N=10000: ${\color{red} \boxdot}$).
 The shaded areas underneath represent the distributions of coefficients $C_n=\langle E_n\vert \Psi_0\rangle$ of the initial states  
 discussed above (N=1000: blue; N=10000: red;  $p_L(0)$=0.8: distributions on the left ;  $p_L(0)$=1.0: distributions on the right). 
 The distributions are sharply peaked around the index $n=n_{max}$ 
 for which the eigenstate energy $E_n$ equals the energy of the initial state $E$ 
 (vertical lines). The corresponding eigenstate fragmentation $f_{n_{max}}$ is indicated as well. 
 The analytical result  (solid black line) 
 for the eigenstate fragmentation $f_{n}$ as function of the eigenstate energy $E_n$ is also shown and 
 approximates the numerical one well. 
 Comparing the left and right panel it is seen that $f_{col}=f_{n_{max}}$.  
 This connection between the initial state and the fragmentation after the collapse of the density oscillations
 holds numerically for the entire family of initial states $\vert\Psi_0\rangle$ and is
 predicted analytically by Eq. (\ref{fcolfn}).  See text for details. 
 All quantities shown are dimensionless.}
 \label{F3}
 \end{figure}

We consider the family of condensed initial states given by
\begin{equation}\label{psi0}
\vert \Psi_0\rangle=\frac{1}{\sqrt{N!}}\left(\sqrt{p_L(0)} b_L^\dag + 
\sqrt{p_R(0)} b_R^\dag \right)^N\vert 0\rangle
\end{equation}
and begin by choosing parameter values corresponding to the case considered 
in the previous section for the many-body Schr\"odinger dynamics, $\lambda=0.152$ ($\Lambda=1.33$).  For the initial states Eq. (\ref{psi0})
we solve the dynamics using the BH Hamiltonian (\ref{BHHam}).
We find that for the entire family of initial states $\vert\Psi_0\rangle$ with $0\le p_L(0)\le 1$,
BECs consisting of different numbers of particles
fragment to the same value $f_{col}$ during the collapse of the density oscillations.
Thus, also the BH fragmentation dynamics is universal.
This can be seen in Fig. \ref{F3} (left) where the BH fragmentation is shown 
as a function of time for $N=1000$ and $10000$ bosons.
The fragmentation of initial states with $p_L(0)=1.0$ reaches a plateau at about 
$f_{col}=32\%$, not far from the many-body Schr\"odinger result $34\%$, see Fig. \ref{F1} (bottom).
Similarly, initial states with $p_L(0)=0.8$ fragment towards the value $f_{col}=16\%$. 
Thus, the universality of fragmentation is a general phenomenon 
which exists for a whole family of initially-condensed systems. 

\begin{figure}[]
    \centering
    \includegraphics[scale=0.35,angle=-90]{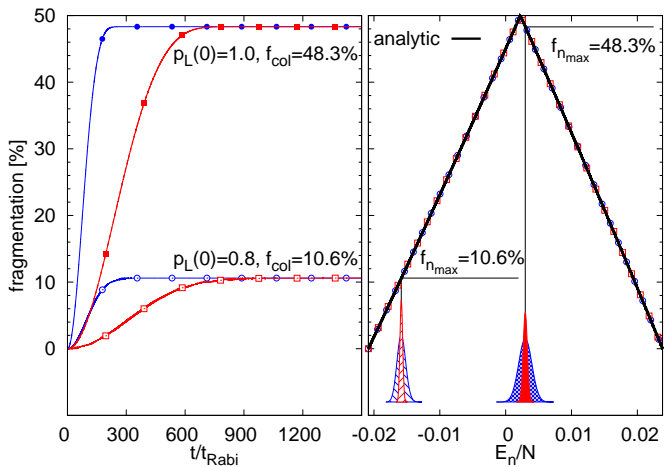}
    	\caption{(color online).  Universality of the Bose-Hubbard fragmentation dynamics for weaker interaction.
 The mean-field interaction strength is now ten-times smaller,  $\lambda=0.0152$ ($\Lambda=0.133$). 
 Left panel: as in Fig. \ref{F3}. The fragmentation after the collapse of the density oscillations is now $f=f_{col}=48.3\%$ for initial states
 with $p_L(0)=1.0$. Here, the approximative formula for $\Lambda\ll 1$, Eq. (\ref{weakfcol}), predicts $f_{col}=1/2-\Lambda/8=48.3\%$ 
 and agrees perfectly with the data. 
 For initial states with $p_L(0)=0.8$ one finds $f=f_{col}=10.6\%$.
 These values are, respectively, higher and lower than for stronger interaction.  
 Note the different, slower time scale compared to the case of stronger interaction.
 Right panel:  as in Fig. \ref{F3}. 
 The analytical result  (solid black line) for the eigenstate fragmentation $f_{n}$ as function of the 
 eigenstate energy $E_n$ provides an excellent approximation to the numerical one here.  
 By comparing the left and right panels one again finds numerically that $f_{col}=f_{n_{max}}$ 
 for the entire family of condensed initial states $\vert\Psi_0\rangle$, 
 as predicted analytically by Eq. (\ref{fcolfn}). 
 All quantities shown are dimensionless.}
    \label{F4}
\end{figure}

Since fragmentation is a phenomenon that occurs due to interparticle interactions, 
we now investigate whether the universality of fragmentation is
robust with respect to reducing the interaction strength. 
For this purpose we consider an interaction strength that is ten times smaller, $\lambda=0.0152$ ($\Lambda=0.133$). 
The results for the fragmentation as a function of time 
are shown in Fig. \ref{F4} (left). 
Analogous to Fig. \ref{F3} the dynamics is shown for the two condensed initial states 
with $p_L(0)=1.0$  and $p_L(0)=0.8$.

As in the case for stronger interaction we find that BECs with different particle numbers fragment 
to the same value in the course of time. 
Condensed initial states with all particles in the left well,  $p_L(0)=1.0$,  
now fragment to about $f_{col}=48.3\%$ which is higher than for stronger interaction. 
Condensed initial states with $p_L(0)=0.8$ fragment to about $f_{col}=10.6\%$ 
which is lower than for stronger interaction.
Note that the time scale is now roughly ten times larger for this interaction strength. 
Nevertheless, even for large particle numbers the BECs eventually fragment to the same value.
Thus, the BH dynamics is universal and many-body in nature over a wide range of interaction strengths.

\subsection{Analytical prediction of the universality of fragmentation}
In this section we provide an analytical explanation of the universality of 
fragmentation based on the BH model. 
To this end we calculate approximatively the elements of the BH first-order reduced-density matrix, Eq. (\ref{RDM}),  after the collapse of the density oscillations, 
diagonalize the matrix to obtain the eigenvalues $n^{(1)}_i$
and calculate the fragmentation.

We consider the family of initial states given by Eq. (\ref{psi0}). 
The energy of the initial state 
$E=\langle\Psi_0\vert H_{BH}\vert \Psi_0\rangle$
can be evaluated and yields
\begin{align}\label{initstateenergy}
E=-2NJ\sqrt{p_L(0)p_R(0)}+\frac{UN(N-1)}{2}(p_L(0)^2+p_R(0)^2). 
\end{align}
Next, we consider the spectrum of the BH Hamiltonian
\begin{equation}
\hat H_{BH}\vert E_n\rangle=E_n\vert E_n\rangle.
\end{equation}
and note that all eigenstates  are parity eigenstates $\hat{P}\vert E_n\rangle=\pm\vert E_n\rangle$ because the potential $V(x)$ is symmetric.
The time-dependent BH wave function 
$\vert \Psi(t)\rangle$ can be expanded in the BH eigenstates $\vert E_n\rangle$ 
\begin{align}
\vert \Psi(t)\rangle =  \sum_{m=0}^N C_n  \vert E_n\rangle e^{-iE_nt}
\end{align}
with coefficients $C_n=\langle E_n\vert\Psi(0)\rangle$. 
By treating the interaction as a perturbation to the noninteracting system  the  first-order perturbative expression for the BH energy eigenvalues reads
\begin{align}\label{eigenenergy}
E_n=-J(N-2n)+\frac{U}{2}(N-n)n &+\frac{UN(N-1)}{4}.
\end{align}
For a noninteracting system the coefficients $C_n$ are binomially distributed with a width $\sim1/\sqrt{N}$, i.e. the distribution becomes sharply peaked for $N\gg1$.  
The energy levels are equidistantly spaced $E_{n+1}-E_n=2J$. 
Thus, in the noninteracting case the system of bosons always returns precisely to its initial state after a time interval of 
$\pi/J=t_{Rabi}$, regardless of its initial state. 
However, a nonzero interaction modifies the energy spectrum, shifts the distribution of the $C_n$ and causes the contributions from different eigenstates to dephase in time.
This leads eventually to the collapse of the density oscillations.

Assuming that the distribution of coefficients $C_n$ also remains sharply peaked  around a value $n_{max}$ in the interacting case, the 
location of the maximum of the distribution can be determined by requiring 
that the energy of the initial state equals the energy of a BH eigenstate. 
By equating Eqs. (\ref{eigenenergy}) and (\ref{initstateenergy}) one finds 
\begin{align}\label{eqnmax}
\frac{n_{max}}{N}&\phantom{+}=\frac{1}{2}+\frac{1}{\Lambda}- \nonumber \\
&\sqrt{\frac{1}{\Lambda^2}+\frac{2\sqrt{p_L(0)p_R(0)}}{\Lambda}+\frac{3}{4}-\left( p_L(0)^2+p_R(0)^2 \right)}
\end{align}
in the limit where $N-1\approx N$. 
Using $\hat b_{L,R}=\frac{1}{\sqrt{2}}(\hat b_{g}\pm\hat b_{u})$ and $\hat n_i=\hat b^\dagger_i\hat b_i$ the elements of the first-order reduced density matrix read
\begin{align}
\rho_{LL}(t)&=\frac{1}{2} \langle\Psi(t)\vert  \hat{n}_g +\hat{n}_u + \hat b_u^\dagger \hat b_g  +\hat b_g^\dagger \hat b_u  \vert \Psi(t)\rangle \label{rhoLL}\\
\rho_{RR}(t)&=\frac{1}{2}\langle\Psi(t)\vert  \hat{n}_g +\hat{n}_u - \hat b_u^\dagger \hat b_g  -\hat b_g^\dagger \hat b_u  \vert \Psi(t)\rangle \label{rhoRR} \\
\rho_{LR}(t)&=\frac{1}{2} \langle\Psi(t)\vert  \hat{n}_g -\hat{n}_u + \hat b_u^\dagger \hat b_g  -\hat b_g^\dagger \hat b_u \vert \Psi(t)\rangle\label{rhoLR}.
\end{align}
Assuming that oscillatory terms average to zero and using the fact that due to symmetry 
$\langle E_n\vert \hat{b}_u^\dagger \hat{b}_g  \vert E_n\rangle = \langle E_n\vert \hat{P}^\dagger\hat{P}  \hat{b}_u^\dagger \hat{b}_g  \hat{P}^\dagger\hat{P}\vert E_n\rangle=-\langle E_n\vert \hat{b}_u^\dagger \hat{b}_g  \vert E_n\rangle=0$
one finds
\begin{align}
\langle \Psi(t)\vert  \hat{b}_u^\dagger \hat{b}_g  \vert \Psi(t)\rangle =  \qquad \qquad \qquad \qquad\qquad \qquad  \nonumber\\\sum_{m\neq n} C_m^\ast C_n \langle E_m\vert   \hat{b}_u^\dagger \hat{b}_g  \vert E_n\rangle e^{i(E_m-E_n)t}=0,  \label{ngu1}
\end{align}
and similarly 
\begin{align}
\langle \Psi(t)\vert & \hat{n}_{u}  \vert \Psi(t)\rangle     \nonumber\\ 
&=\sum_{m,n}  C_m^\ast C_n  \langle E_m\vert   \hat{n}_{u} \vert E_n\rangle e^{i(E_m-E_n)t}  \nonumber\\ 
&=\sum_{n}  |C_n|^2 \langle E_n\vert   \hat{n}_{u} \vert E_n\rangle \approx n_{max}  \label{ngu2}
\end{align}
where in the last approximation it was used that the coefficient distribution is sharply peaked at $n=n_{max}$. 
Substituting Eqs. (\ref{ngu1}) and (\ref{ngu2}) into Eqs. (\ref{rhoLL}), (\ref{rhoRR}) and (\ref{rhoLR}) and using $N=n_u+n_g$ 
one arrives at 
\begin{align}
\rho_{LL}&=\rho_{RR}=\frac{N}{2}\\
\rho_{LR}&=\frac{N}{2}-n_{max}.
\end{align}
By diagonalizing the matrix $\rho_{ij}$ one finds $n^{(1)}_{1,2}=n_{max},N-n_{max}$ and thus using Eq. (\ref{defFrag}) 
\begin{equation}\label{fcol}
f_{col}=\min\left[\frac{n_{max}}{N},1-\frac{n_{max}}{N}\right].
\end{equation}
The universality of fragmentation for different particle numbers now follows directly from Eq. (\ref{eqnmax}):
the values for $n_{max}/N$ and therefore for $f_{col}$ depend on $N$ only through the parameter $\Lambda$
and thus all systems with the same value of $\Lambda$ fragment to the same value $f_{col}$, regardless of the number of particles. 
We also note that  the fragmentation of each eigenstate $\vert E_n\rangle$ is given  
to lowest order by 
\begin{equation}\label{fnanalytic}
f_{n}=\min\left [\frac{n}{N},1-\frac{n}{N}\right].
\end{equation} 
Therefore the fragmentation after the collapse of the density oscillations $f_{col}$ given in Eq. (\ref{fcol}) 
is the same as that of the eigenstate around which the distribution of coefficients $C_n$ is peaked,  and thus 
\begin{equation}\label{fcolfn}
f_{col}=f_{n_{max}}.
\end{equation}

It is instructive to  expand Eq. (\ref{eqnmax}) for two limiting cases. 
If initially all particles are in one of the wells,  $p_L(0)=1$ or $p_L(0)=0$, and $\Lambda\ll 1$
Eq. (\ref{fcol}) reduces to
\begin{equation}\label{weakfcol}
f_{col} = \frac{1}{2}-\frac{\Lambda}{8}, 
\end{equation}
predicting strong fragmentation at arbitrarily weak interaction strengths and for arbitrarily large numbers of particles
after the collapse of the density oscillations.

In contrast, for the initial state with equal numbers of particles in both wells, $p_L(0)=p_R(0)=1/2$,  
Eq. (\ref{eqnmax}) predicts $n_{max}=0$ and according to Eq. (\ref{fcol}) 
\begin{equation}\label{pl0505}
f_{col} = 0,
\end{equation}
indicating that no fragmentation should occur. Thus,  whether or not the system of bosons 
fragments is expected to depend strongly on the initial state.

\subsection{Numerical verification}
In this section we test the analytical predictions of the previous section. 
To this end we diagonalize the BH Hamiltonian, Eq. (\ref{BHHam}), and evaluate the respective quantities. 
The right panel of Fig. \ref{F3} shows the fragmentation $f_n$ of each BH eigenstate as a function of its energy per particle 
for $N=1000$ and $N=10000$ bosons. 
The curves for different particle numbers lie on top of each other. 
Also shown are the sharply peaked  distributions of the expansion coefficients $C_n$ 
for the initial states with $p_L(0)=0.8$ and $p_L(0)=1.0$. 
The values of the eigenstate fragmentations $f_n$ at the peaks of the distributions are shown as well.
The solid black line represents the approximative analytical result for the eigenstate fragmentations as a function of the perturbative 
energy per particle, i.e. $f_n$ given by Eq. (\ref{fnanalytic}) as a function
of $E_n/N$ with $E_n$ given by Eq. (\ref{eigenenergy}). The analytic result provides good approximation to the numerical one 
at this interaction strength.
By comparing the left and the right panels of Fig. \ref{F3} 
it is clearly seen that $f_{col}=f_{n_{max}}$, as expected on the basis of Eq. (\ref{fcolfn}).

The same is true for the case of ten-times  weaker interaction strength, shown in Fig. \ref{F4}.  
Comparing the analytical predictions in Figs. \ref{F3} and \ref{F4} it is clearly seen 
that for weaker interaction strength the approximative analytical results improve. 

Lastly, we check the predictions of Eqs. (\ref{weakfcol}) and (\ref{pl0505}). 
For the initial states with $p_L(0)=1$ Eq. (\ref{weakfcol}) predicts a 
fragmentation  after the collapse of $f_{col}=33.3\%$ ($48.3\%$) for $\Lambda=1.33$ $(0.133)$
in good agreement with the numerical values. We note that for the initial state with $p_L(0)=0.5$
there are no density oscillations between the two wells 
and thus there can also be no collapse.
Nevertheless, in agreement with Eq. (\ref{pl0505}) no appreciable fragmentation occurs in the dynamics:
numerically, one finds that the fragmentation of the BH model oscillates between $0\%$ and $0.2\%$. 

These results show that the fragmentation after the collapse of the density oscillations $f_{col}$ depends crucially on the 
initial state:  initial states with large differences in particle numbers 
between the left and the right well fragment strongly, even if $\Lambda\ll1$, i.e. if the mean-field interaction strength $\lambda$ is very small. 
Therefore, there is no weakly interacting limit where the GP mean-field is valid at long times.
On the other hand, condensed initial states with similar numbers of particles 
in both wells do not fragment strongly in time.

\section{Conclusions}
Summarizing, we have shown that there is a universal fragmentation 
dynamics in bosonic Josephson junctions at weak interaction strengths by solving the time-dependent
many-boson Schr\"odinger equation. The phenomenon can be detected, for instance, 
through a measurement of the two-particle momentum correlation functions.
We have shown that the two-mode Bose-Hubbard model  approximates
the many-body Schr\"odinger results quite well here.  
The origin of the universal fragmentation dynamics 
can be explained analytically and numerically using the two-mode Bose-Hubbard model.
We found that the extent to which many-body effects become important 
at later times depends crucially on the initial state. 
The value of the fragmentation of initially condensed  states  at later times 
can be precisely predicted analytically in the limit of weak interactions. 
At arbitrarily weak interaction strength 
and for arbitrarily large numbers of particles there are condensed initial states that fragment strongly.
Thus, there is no weakly interacting limit where the Gross-Pitaevskii mean-field is valid at long times.

\begin{acknowledgments}
Financial support by the DFG is gratefully acknowledged  as is computing time at the High Performance Computing Center Stuttgart (HLRS). 
K.S. acknowledges funding through the Karel Urbanek Postdoctoral Research Fellowship. Discussions with Mark Kasevich are gratefully acknowledged.
\end{acknowledgments}

\appendix
\section{Logarithmic scaling of the fragmentation time}\label{fragtime}
 \begin{figure}[t]
 \includegraphics[scale=0.35,angle=-90]{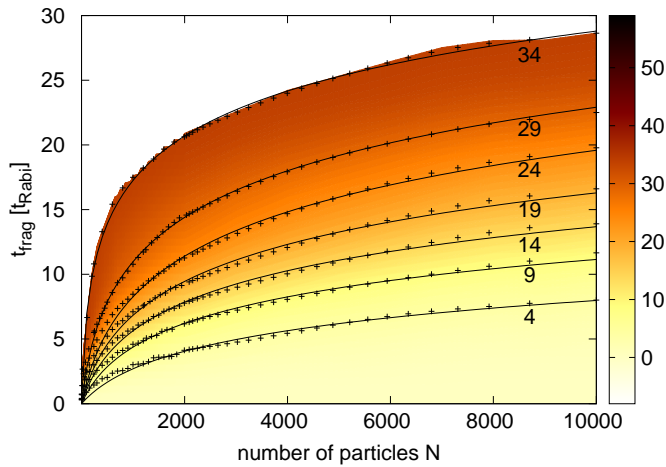}
 \caption{(color online): Many-body Schr\"odinger fragmentation times. Shown are the times $t_{frag}$ that 
 are needed to reach a given fragmentation $f$ as a function of the number of bosons for the same 
 parameters as in Fig. 1. The fragmentation values (crosses) are well described by a
 logarithmic fit function (lines), here shown for the values $f=4\%,9\%,\dots,34\%$. 
 This logarithmic dependence of $t_{frag}$ on the number of particles implies the 
 breakdown of GP mean-field theory even in the limit of large particle numbers.
 See text for details. All quantities shown are dimensionless.}
 \label{F5}
 \end{figure}

Here we briefly discuss the time scales involved in the Schr\"odinger fragmentation dynamics shown in Fig. \ref{F1}.
We define the fragmentation time $t_{frag}$ as the first time 
at which a certain fragmentation $f$ is reached.
Fig. \ref{F5}  shows $t_{frag}$ 
as a function of $N$. Clearly, $t_{frag}$ increases with $N$.
For any value of $f$, $t_{frag}$ 
is well described by a fit to the function $t(N)=a \ln({1+bN})$, as is shown here for 
the values $f=4\%,9\%,...,34\%$. Thus, $t_{frag}$ grows logarithmically 
with $N$. Consequently, the fragmentation 
does not decrease or even disappear in the limit of large $N$. 
Even for $N=10000$ bosons the fragmentation 
rises to about $10\%$ in less than a dozen $t_{Rabi}$.
As GP theory does not allow BECs to fragment at all, 
$t_{frag}$ defines a measure for the breakdown of mean-field theory.
Thus, we find here a  logarithmic dependence of the mean-field theory  breakdown time 
on $N$ based on the time-dependent many-body Schr\"odinger dynamics. 
We note that a logarithmic breakdown of mean-field theory has been reported earlier based on the BH model \cite{Vardi01}.

\end{document}